\documentclass[aps,prb,reprint]{revtex4-2}%
\usepackage{amsmath}
\usepackage{array}
\usepackage{array}
\usepackage{amsfonts}
\usepackage{amssymb}
\usepackage{graphicx}
\usepackage{color}
\usepackage{dcolumn}
\usepackage{fullpage}
\usepackage{bm}
\usepackage
{hyperref}
\usepackage{epstopdf}
\usepackage{cancel}

\begin{document}

\title{Dzyaloshinskii-Moriya spin density by skew scattering}
\author{\href{https://orcid.org/0000-0002-2068-9613}{Adam B. Cahaya}$^1$ and \href{https://orcid.org/0000-0001-9078-685X}{Alejandro O. Leon}$^2$}
\affiliation{$^1$ Department of Physics, Faculty of Mathematics and Natural Sciences, Universitas Indonesia, Depok 16424, Indonesia\\
$^2$ Departamento de F\'isica, Facultad de Ciencias Naturales, Matem\'atica y del Medio Ambiente, Universidad Tecnol\'ogica Metropolitana, Las Palmeras 3360, Ñuñoa 780-0003, Santiago, Chile}

\begin{abstract}
\begin{small}
Anisotropic exchange couplings, such as the Dzyaloshinskii-Moriya interaction (DMI), have played a vital role in the formation and dynamics of spin textures. This work predicts {an anisotropic} conduction electron spin density in metals with heavy magnetic impurities. The polarization of this \textit{Dzyaloshinskii-Moriya spin density} (DM-SD) is not collinear to the localized magnetic moments but rotated by the spin-dependent skew scattering of heavy atoms. The DM-SD induces the DMI between magnetic moments in metals and, therefore, it is the anisotropic extension of the Rutherman-Kittel-Kasuya-Yoshida spin density. Our model consists of two localized magnetic moments, one with a large spin-orbit coupling (a lanthanide or rare earth), in a free electron gas. The lanthanide spin controls the DM-SD strength and polarization, promising a flexible control mechanism for anisotropic couplings. 
\end{small}
\end{abstract}
\maketitle 

\begin{small}
\textit{Introduction.} 
Characterizing the exchange interaction~\cite{BookSkomski1,BookBlundell} has been fundamental for understanding magnetic properties and textures. For example, the Dzyaloshinskii-Moriya interaction (DMI), a type of anisotropic exchange coupling, has been extensively studied for its capacity to tune magnetic characteristics~\cite{BookBlundell,BookHandBook1,BookHandBook3} and stabilize topological spin textures, such as skyrmions~\cite{BookHandBook3,SkyrmionArticles1,SkyrmionArticles2,SkyrmionArticles3} and merons~\cite{EzawaArticle}.
Historically, the proposal of an anisotropic exchange to explain the emergence of magnetization in certain antiferromagnets by Dzyaloshinskii \cite{BookMagnetismTMO,DzyaloshinskiiPaper} was followed by Moriya's derivation of the DMI Hamiltonian~\cite{BookMagnetismTMO,MoriyaPaper} from Anderson's superexchange theory~\cite{Andersons}.
Fert and Levy~\cite{DMIMnPtFert,DMIMnPtFert2} created a model of two magnetic atoms in a normal metal in the presence of a nonmagnetic impurity with spin-orbit coupling (SOC). Subsequently, in a model based on virtual-bound states~\cite{VBS} where Fe ions provide both the exchange and the SOC, the DMI energy~\cite{DMISOC1} was found to depend on the relative directions and distances of the spins of the three atoms. In rare earths in noble metals, a DMI energy expression was derived~\cite{DMISOC2} from the spin-spin coupling between impurities and conduction electrons and a fine-structure-type SOC. This rare-earth DMI energy~\cite{DMISOC2} is fourth order in the spin-spin exchange and first order on the SOC parameters.  
The applications of Fert and Levy's model~\cite{DMIMnPtFert,DMIMnPtFert2} have been extended to interfaces~\cite{DMIInterfaces1,DMIInterfaces2,DMIInterfaces3,DMIInterfaces4,DMIMultilayers1,DMIMultilayers2} where the inversion symmetry is naturally broken. 
Studies of the DMI in systems with Rashba~\cite{DMIRashba1,DMIRashba2,DMIRashba3,DMIRashba4,DMIRashba5,DMIRashba6,DMIRashba7,DMIRashba8} and Dresselhaus~\cite{DMIRashba2,DMIRashba8,DMIRashba9} SOCs and ferroelectric materials~\cite{DMIFerroelectric1,DMIFerroelectric2} have been conducted. In addition,  the control of the DMI by electric fields, interface coverage, and currents~\cite{DMIOxygen,DMITuning,DMICurrent,DMITuning2,DMICurrent2}, as well as its origin in terms of first-principles calculations~\cite{DMIDFT} and equilibrium spin currents and Berry curvature~\cite{DMISpinCurrent,DMIBerryCurvature,DMISpinCurrentReview}, and orbital-momentum anisotropy~\cite{DMIOrbitalMom}, have also gathered substantial attention.

Here, we study the conduction-electron spin density around two magnetic atoms, one of them with a large SOC. The polarization of this conduction-electron \textit{Dzyaloshinskii-Moriya spin density} (DM-SD) is not collinear to the localized magnetic moments but has a mutually orthogonal component (see Fig.~\ref{Fig1}). Furthermore, the DMI energy emerges naturally when a third atom locally interacts with the DM-SD.
Therefore, the DM-SD is responsible for the DMI in the same way the {Rutherman-Kittel-Kasuya-Yoshida (RKKY)~\cite{RKKY-1,RKKY-2,RKKY-3}} spin density mediates the isotropic exchange between localized spins. In our model, the metal is a free electron gas, and the SOC source is a lanthanide atom; they interact via the spin-dependent skew scattering of the Kondo Hamiltonian~\cite{Kondo} that deflects in opposite directions electrons with different spins. Therefore, the coupling between conduction electrons' trajectories and spins,\textit{ i.e.,} the effective SOC, is well localized around the lanthanide atom, while the rest of the metal is SOC-free.

\begin{figure}[t!]
\centering\includegraphics[width=\columnwidth]{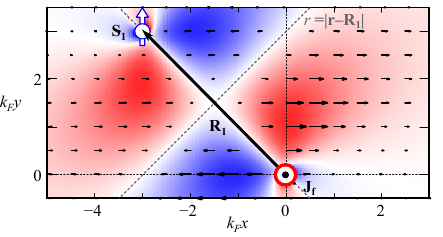}
\caption{\footnotesize Arrow representation of the \textit{Dzyaloshinskii-Moriya spin density} (DM-SD). The polarization of the DM-SD (small black arrows) is orthogonal to the one of the spin sources $\textbf{S}_1$ (at $\mathbf{R_1}$) and $\mathbf{S_f}$ (at the origin). The latter has spin-orbit coupling mediated by the lanthanide orbital momentum $\mathbf{L_f}$. In this figure $\mathbf{L_f}\parallel\mathbf{S_f}\parallel\mathbf{J_f}\equiv\mathbf{L_f}+\mathbf{S_f}$, $k_F$ is the Fermi wavenumber of the host metal and $x$ and $y$ are the components of the position vector $\mathbf{r}$.}
\label{Fig1}
\end{figure}

\textit{Rare-earth impurities in a simple metal.} Rare-earth (RE) atoms are Sc, Y, and lanthanides. They are valuable spintronic components that tune the magnetic properties in insulators~\cite{REsGeneralities1,REsGeneralities2,REsGeneralities3,REsGeneralities4,REsGeneralities5,REsGeneralities6}. Lanthanides with large orbital momentum enrich the phenomenology of metallic devices~\cite{REsGeneralitiesI1,REsGeneralitiesI2,REsGeneralitiesI3}. 
Furthermore, interfacial DMI at RE-containing magnetic-insulator$\vert$heavy-metal bilayers has been the subject of intense research in the past few years~\cite{DMITIG1,DMITIG2,DMITIG3,DMITIG4,DMITIG5}. Here we focus on the tripositive RE ions: Ce$^{3+}$, Pr$^{3+}$, Nd$^{3+}$, Pm$^{3+}$, Sm$^{3+}$, Tb$^{3+}$, Dy$^{3+}$, Ho$^{3+}$, Er$^{3+}$, Tm$^{3+}$, and Yb$^{3+}$, that possess a substantial orbital momentum $\mathbf{L}_{\rm f}$ that strongly couples to the RE spin $\mathbf{S}_{\rm f}$, forming a net angular momentum $\mathbf{J}_{\rm f}=\mathbf{L}_{\rm f}+\mathbf{S}_{\rm f}$. 
The origin of such momenta is the well-localized and partially filled 4f subshell \cite{BookSkomski2}. The SOC of 4f electrons mediates the interaction between electronic, mechanical, and magnetic degrees of freedom in lanthanide-based materials~\cite{BookSkomski1,BookBlundell,REsGeneralities5}. We model the REs' spin [$\mathbf{S}_{\rm f}=(g_J-1)\mathbf{J}_{\rm f}$] and orbital [$\mathbf{L}_{\rm f}=(2-g_J)\mathbf{J}_{\rm f}$] momenta, with $g_J$ being the Land\'e g-factor, as classical vectors with well-defined norms embedded in a free electron gas. The latter is described by plane waves, $\psi_{\mathbf{k}}(\mathbf{r})=e^{i\mathbf{k}\cdot\mathbf{r}}/\sqrt{V_0}$, where $\hbar\mathbf{k}$ is the linear momentum, $\hbar$ is the reduced Plank constant, and $V_0$ is the system volume. The magnetic part of the Kondo Hamiltonian $H_{\rm kondo}=H_{Ss}+H_{Ll}+H_{SsLl}$ accounts for the interactions between the RE and conduction electrons~\cite{Kondo}. The {$H_{Ss}\propto\mathbf{S}_{\rm f}\cdot\boldsymbol{\sigma}$} term is the spin-spin interaction with exchange constant $J_{\rm 4f}$ (similar to the $J_{\rm 3d}$ parameter of transition-metal ions), { where $\boldsymbol{\sigma}$ is the vector of Pauli matrices. This interaction produces a collinear, or RKKY~{\cite{RKKY-1,RKKY-2,RKKY-3}}, spin polarization near the impurity}. Next, the spin-independent skew scattering {$H_{Ll}\propto\mathbf{L}_{\rm f}\cdot\left[\mathbf{k}_1\times\mathbf{k}_2\right]$ generates anomalous-Hall effect and} an equilibrium rotational current~\cite{REsGeneralitiesI3}. Finally, $H_{SsLl}$ is the spin-dependent skew scattering that deflects in opposite directions up and down spins, as illustrated by its matrix elements,
\[\left\langle \textbf{k}_1\alpha\right\vert H_{SsLl}\left\vert \textbf{k}_2\beta\right\rangle
= -\frac{i\xi}{V_0}\left(\boldsymbol{\sigma}_{\alpha\beta}\cdot\textbf{S}_{\rm f} \right)\left(\textbf{L}_{\rm f}\cdot \left[\textbf{k}_1\times\textbf{k}_2\right]\right),\]
where $\alpha$ and $\beta$ are spin labels, and $\xi$ is the interaction constant. The energy of the spin-dependent and spin-independent~\cite{REsGeneralitiesI3} skew scatterings are similar~\cite{Kondo}, with $\xi\hbar^2k_F^2\sim 0.1$~eV\AA$^3$ for an Al host metal. {The scales of $H_{Ll}$ and $H_{SsLl}$ are at least one order of magnitude smaller than the one of $H_{Ss}$. Therefore, our theory will neglect contributions non-linear in $\xi$.}

The combined effects of the spin-spin exchange and the SOC give rise to an anisotropic polarization in the conduction electron spin density, as shown in the following paragraphs.
 
 \textit{Dzyaloshinskii-Moriya spin density.} Let us consider two magnetic ions, one is a transition metal (without SOC) of spin $\mathbf{S_1}$ located at $\mathbf{R_1}$ and the other is a RE with momenta $\mathbf{S}_{\rm f}$, $\mathbf{L}_{\rm f}$, and $\mathbf{J}_{\rm f}$ at the origin, $\mathbf{r}=0$. If $H_{Ll}=H_{SsLl}=0$, then the ensemble expectation value of the conduction-electron spin density $\mathbf{s}\left(\mathbf{r}\right)$, $ \langle \mathbf{s}\left(\mathbf{r}\right)\rangle={\rm Tr}[\rho\mathbf{s}\left(\mathbf{r}\right)]$, with $\rho$ being the density matrix and Tr the trace, reads
 \begin{align}
 \langle \mathbf{s}\left(\mathbf{r}\right)\rangle\approx
 J_\text{3d}   \chi_0\left(\textbf{r}-\textbf{R}_1\right)  \textbf{S}_1+J_\text{4f}   \chi_0\left(\textbf{r}\right)  \textbf{S}_f,
 \label{MeanValueS}
 \end{align}
 where the RKKY response function is
 \begin{align}
\chi_0(\mathbf{r})= \frac{\text{DOS}(k_F)}{4\pi r^3}\left(\frac{\sin 2k_Fr}{2k_Fr}-\cos 2k_Fr\right),
\label{RKKYSimple}
\end{align}
and $r=\vert\mathbf{r}\vert$, the density of states is DOS$(k_F)=m_ek_F/\left(\pi\hbar\right)^{2}$, $m_e$ is the electron mass, and $k_F$ is the Fermi wavenumber. It is worth noting that the approximation of Eq.~(\ref{MeanValueS}) retains terms {proportional to} $J_\text{3d} $ or $J_\text{4f}$, as given by the Kubo formula. If, on the other hand, $H_{SsLl}\neq0$ and terms of order $J_\text{3d}\xi$ are admitted, the Dzyaloshinskii-Moriya spin density (DM-SD), $ \langle \mathbf{s}\left(\mathbf{r}\right)\rangle_{\rm {DM}}$,  emerges and is added to Eq.~(\ref{MeanValueS}), as shown in the Supplemental Material~\cite{SupplMat}
\begin{align}
 \langle \mathbf{s}\left(\mathbf{r}\right)\rangle_{\rm {DM}}=&
 -f(r,R_1,r_1)\textbf{S}_{\rm f}\times\textbf{S}_1\mathbf{L_f}\cdot\left(\hat{\textbf{R}}_1\times\hat{\textbf{r}}\right),\notag\\
  f(r,R_1,r_1)=&\frac{\xi J_{\rm 3d} m_e^2}{\left(2\pi \right)^4\hbar^2} \frac{F \left(R_1,r,r_1\right)}{R_1rr_1},
  \label{EqMainResults}
\end{align}
 where $\hat{\mathbf{r}}=\mathbf{r}/r$, $r_1=\vert\mathbf{r}-\mathbf{R_1}\vert$, and
 \begin{figure}[t!]
\centering\includegraphics[width=\columnwidth]{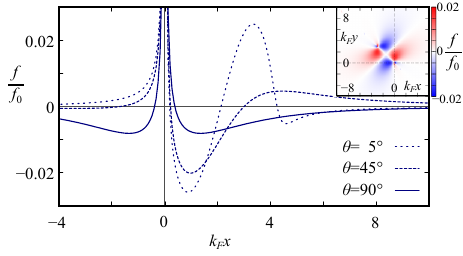}
\caption{\footnotesize Function $f$, obtained from Eq.~(\ref{EqMainResults}), along the $x$ axis for several angles $\theta$ between $\mathbf{R_1}$ and $\mathbf{r}$ (with the RE at the origin). The normalization is $f_0=\xi J_{\rm 3d}m_e^2k_F^4\hbar^{-2}\left(2\pi\right)^{-4}$. The spatial dependence of the DM-SD, $ \langle \mathbf{s}\left(\mathbf{r}\right)\rangle_{\rm {DM}}$, is given by $\hat{\textbf{R}}_1\times\hat{\textbf{r}}$ and the $f$ profile. The inset shows the spatial extent of the DM-SD {in the $x$-$y$ plane}. }
\label{Fig2}
\end{figure}%
\begin{align*}
&F\left(R_1,r,{r_1}\right)\\&=
-\frac{3 \left(k_F^2 (r-r_1+R_1)^2-2\right) \sin (k_F (r-r_1+R_1))}{(r-r_1+R_1)^4}\\&
-\frac{3 \left(k_F^2 (-r+r_1+R_1)^2-2\right) \sin (k_F (r-r_1-R_1))}{(-r+r_1+R_1)^4}\\&+\frac{k_F \left(k_F^2 (-r+r_1+R_1)^2-6\right) \cos (k_F (r-r_1-R_1))}{(r-r_1-R_1)^3}\\&
+\frac{k_F \left(k_F^2 (r-r_1+R_1)^2-6\right) \cos (k_F (r-r_1+R_1))}{(r-r_1+R_1)^3}.
\end{align*}
Figure~\ref{Fig2} exemplifies the spatial dependence of the DM-SD. This type of spatial map may be a valuable tool for engineering interfaces with large DMI. Note that the spin density at $\mathbf{r}$ vanishes if $\left(\textbf{R}_1\times\textbf{r}\right)\perp\mathbf{L_f}$ or $r=\vert\mathbf{r}-\mathbf{R_1}\vert$, {as illustrated by Fig.~\ref{Fig1}}. In addition, $\langle \mathbf{s}\left(\mathbf{r}\right)\rangle_{\rm {DM}}\propto\textbf{S}_{\rm f}\times\textbf{S}_1$ and then the RE angular momentum rules the strength and orientation of the DM-SD.

Within our model, the relative importance of the DM-SD and the RKKY spin density is given by the ratio $\xi\hbar^2k_F^5/E_F\sim10^{-1}$, with $E_F=\hbar^2k_F^2/\left(2m_e\right)$ being the Fermi energy of the host metal.

\begin{figure}[t]
\centering\includegraphics[width=\columnwidth]{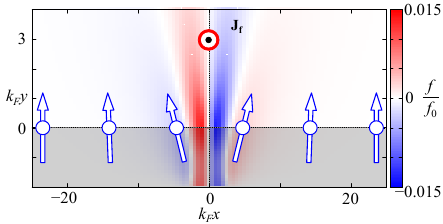}
\caption{\footnotesize Spin chain in the presence of the Dzyaloshinskii-Moriya spin density (DM-SD). Atomic spins locally interact with the DM-SD (background color around the rare-earth angular momentum $\mathbf{J}_{\rm f}$), which favors a tilting from the fully parallel state. The orientation of the magnetic moments is the result of the competition between the RKKY spin density and the DM-SD, as well as other interactions, including anisotropies. In this figure, the tilting is exaggerated in the $x-$axis for pedagogical reasons. The zones above (white) and below (grey) the spin chain could be interpreted as different materials.}
\label{Fig3}
\end{figure}%

The \textit{s-d} exchange coupling of another transition-metal impurity of spin $\mathbf{S_2}$ at $\mathbf{R_2}$ to $ \langle \mathbf{s}\left(\mathbf{r}\right)\rangle_{\rm {DM}}$ (see the Supplemental Material~\cite{SupplMat}) is
\begin{align}
H_{\rm DMI}&=-J_{\rm 3d} \mathbf{S_2}\cdot\langle \mathbf{s}\left(\mathbf{R_2}\right)\rangle_{\rm {DM}}=\textbf{D}^f_{12}\cdot\left(\textbf{S}_{1}\times\textbf{S}_{2}\right),\notag\\
\textbf{D}^f_{12}
=&\bar{f}(R_1,R_2,\vert\mathbf{R_1}-\mathbf{R_2}\vert)\textbf{S}_f \left[\mathbf{L}_{\rm f}\cdot\left(\hat{\textbf{R}}_1\times\hat{\textbf{R}}_2\right) \right]
\label{EqFinalResult}
\end{align}
with $\bar{f}(R_1,R_2,r_{12})=J_{\rm 3d}\textbf{S}_{\rm f}\big[ f(R_1,R_2,r_{12})+f(R_2,R_1,$ $r_{12}) \big]$ and $r_{12}=\vert\mathbf{R_2}-\mathbf{R_1}\vert$. The above expressions provide a simple interpretation of the DMI that goes beyond the RE physics, \textit{i.e.}, an atomic spin-orbit coupling rotates the RKKY spin density. This tilted spin density (see Figs.~\ref{Fig1} and~\ref{Fig2}), or Dzyaloshinskii-Moriya spin density, couples to other local moments via the \textit{s-d} or \textit{s-f} exchange interaction, giving rise to an anisotropic exchange. Therefore, the spin density is the entity that mediates the DMI. This intuitive picture is similar to the RKKY spin polarization resulting in the exchange between two magnetic particles. { The DMI favors non-collinear textures, as exemplified by Fig.~\ref{Fig3}, where a DM-SD tilts the orientation of local magnetic moments in an anisotropic fashion. In this figure, the magnetic moments (arrows) point towards the local spin density.}

In the model by Fert and Levy~\cite{DMIMnPtFert,DMIMnPtFert2}, the DMI energy has the form $\mathbf{R_1}\cdot\mathbf{R_2}\left(\mathbf{R_1}\times\mathbf{R_2}\right)\cdot\left(\mathbf{S_1}\times\mathbf{S_2}\right)$. The $\mathbf{R_1}\cdot\mathbf{R_2}$ dot product arises from using virtual-bound states around the heavy impurity~\cite{VBS}. Therefore, the DMI strength,  mediated by virtual-bound states, is proportional to $\sin(2\theta)$, where $\theta$ is the angle between $\mathbf{R_1}$ and $\mathbf{R_2}$. Thus, this DMI expression vanishes if $\mathbf{R_1}\perp\mathbf{R_2}$. In the study of Fe or REs in noble metals~\cite{DMISOC1,DMISOC2}, an additional energy term proportional to $\mathbf{R_1}\cdot\mathbf{R_2}\left(\mathbf{R_1}\times\mathbf{R_2}\right)\cdot\mathbf{S_3}\left(\mathbf{S_1}\times\mathbf{S_2}\right)\cdot\mathbf{S_3}$ was found, where $\mathbf{S_{3}}$ is the spin of the third atom.
Meanwhile the DMI energy, in Eq.~(\ref{EqFinalResult}), has some similarities with previous results~\cite{DMISOC2}, including the order of magnitude, our work aims to predict and characterize the formation of a conduction-electron spin density that couples magnetic particles. {As it occurs with the RKKY spin density, the DM-SD will modify the spin and charge flow at interfaces or mediate interactions between different magnetic media. In our model}, the spin-dependent skew scattering of the Kondo Hamiltonian is the cause of the tilted spin density $\langle \mathbf{s}\left(\mathbf{R_2}\right)\rangle_{\rm {DM}}$, or DM-SD,  that locally interacts with a third local magnetic moment. This contact skew interaction does not require the formation of virtual-bound states. Therefore, the DMI energy is of third order in the perturbation Hamiltonian ($\sim J_{\rm 3d}^2\xi$) instead of the fifth-order expression ($\sim J_{\rm 3d}^4\lambda$, where $\lambda$ is the SOC parameter). 

Furthermore, since our results do not depend on the formation of the virtual-bound state, the DMI strength {is proportional to $\sin(\theta)$ and} is finite when the positions of the interacting magnetic moments are perpendicular, $\mathbf{R}_1\perp\mathbf{R}_2$, with the SOC source being the center of coordinates. { To illustrate this difference,}
{let us consider a lanthanide at the origin with $\mathbf{S_f}=\mathbf{L_f}=3\hbar\mathbf{e_z}$, and two light magnetic atoms at $\mathbf{R}_1=a\mathbf{e_x}$, $\mathbf{R}_2=a\mathbf{e_y}$, with $a=5$\AA, and spins $\mathbf{S_1}=5\hbar\mathbf{e_x}$ and $\mathbf{S_2}=5\hbar\mathbf{e_y}$. Within Fert and Levy theory, there is no DMI in this system because $\mathbf{R}_1\cdot\mathbf{R}_2=0$. However, in our case the DMI is finite and can be estimated as $H_{\rm DMI}\sim10^{-22}$ J, or, in terms of the DMI density~\cite{ArticleDani}, $D\equiv 2H_{\rm DMI} a^{-2}\sim1$~mJ/m$^{2}$ for $J_{3d}=50$~eV\AA$^3\hbar^{-2}$ and $k_F=1.75$~\AA$^{-1}$. Finally, it} is worth noting that in the DMI of magnetic insulators~\cite{DMIInsulatorsMoskvin1,DMIInsulatorsMoskvin2} {originates on a different mechanism than the one of metals. However, in magnetic insulators}, the DMI vector has the Fert-Levy term $\propto\sin(2\theta)$, as well as the one found here, $\propto\sin(\theta)$.  The exact determination of the relative importance of the two terms in metallic systems requires \textit{ab initio} characterization.

\textit{ Concluding remarks.} In this work we have studied the formation of a spin density around two magnetic atoms, one a transition metal and another a {lanthanide or} rare earth (RE). The spin density has a component parallel to the magnetic moments and another perpendicular to them. {We name the latter Dzyaloshinskii-Moriya spin density (DM-SD).} When this spin density couples with a third magnetic atom, the Dzyaloshinskii-Moriya interaction (DMI) emerges. The physical origin of our predictions is the spin-dependent skew scattering of the Kondo Hamiltonian. This mechanism requires no virtual-bound state and therefore has a different parametric dependence on the positions of the magnetic atoms and the exchange and spin-orbit coupling constants. 

While we have considered a RE atom as the spin-orbit coupling source, the DM-SD also emerges in systems with heavy non-magnetic atoms, such as Pt. The advantage of using REs is their large orbital angular momentum that offers the possibility of controlling the DM-SD by reorienting RE's momenta in, e.g., bilayers composed of a transition-metal-based and a RE-based magnetic materials. We expect this work to open new avenues to understand magnetic interactions and have applications in spintronics devices and topological texture properties.

\textit{Acknowledgments.} We thank Daniela Mancilla-Almonacid for her valuable input on this work's initial calculations and  Roberto E. Troncoso, Alvaro S. Nunez, and Gerrit E. W. Bauer for fruitful discussions. This research was supported by Universitas Indonesia Grant of PUTI NKB-634/UN2.RST/HKP.05.00/2022 and FONDECYT (CL) Grant No. 1210353.
\end{small}

\onecolumngrid 
\newpage

\begin{center}
\begin{large}
\textbf{Supplemental Material to "Dzyaloshinskii-Moriya Spin Density by skew scattering"\\\vspace{1pt}}
\end{large}

{Adam B. Cahaya$^1$ and Alejandro O. Leon$^2$}\\

\textit{$^1$ Department of Physics, Faculty of Mathematics and Natural Sciences,\\ 
Universitas Indonesia, Depok 16424, Indonesia \\
$^2$ Departamento de F\'isica, Facultad de Ciencias Naturales,\\
Matem\'atica y del Medio Ambiente,
Universidad Tecnol\'ogica Metropolitana, \\Las Palmeras
3360, Ñuñoa 780-0003, Santiago, Chile}

\end{center}

\setcounter{page}{1}
\section{RKKY interaction as the coupling between the metal spin density and a localized spin}

The indirect exchange interaction between localized spins $\mathbf{S}_1$ and $\mathbf{S}_2$ can be understood as the formation of a conduction-electron spin density $\mathbf{s}(\textbf{r})$ by a localized impurity spin $\mathbf{S}_1$, and the subsequent interaction between $\mathbf{s}$ and $\mathbf{S}_2$. In the first part of this supplemental material, we review these processes. In the next sections, include a third magnetic atom with spin-orbit coupling.

Consider a classical spin  $\textbf{S}_1$ at position $\textbf{r}=\textbf{R}_1$ embedded into a conduction electron sea. The exchange interaction between $\textbf{S}_1$ and the metal spin density $\mathbf{s}(\textbf{r},t)$ is
\begin{align*}
H_{\rm Ss}= -J_\text{3d} \int d^3r \textbf{s}(\textbf{r},t)\cdot \textbf{S}_1(t)\delta (\textbf{r}-\textbf{R}_1),
\end{align*}
where $J_\text{3d}$ is the exchange constant, such that $J_\text{3d}\hbar^{2}$ has units of energy multiplied by volume.
The above Hamiltonian gives rise to the following linear response of the average metal spin density, $\langle\textbf{s}\rangle_{\rm RKKY}$, to $\textbf{S}_1$
\begin{align}
\langle\textbf{s}\rangle_{\rm RKKY}(\textbf{r},t)=& J_\text{3d} \int d^3{r'}dt' \chi(\textbf{r}-\textbf{r}',t-t') \textbf{S}_1(t')\delta(\textbf{r}'-\mathbf{R}_1)= J_\text{3d} \int dt' \chi(\textbf{r}-\textbf{R}_1,t-t') \textbf{S}_1(t'),
\label{EqLinearInt}
\end{align}
where $\chi$ is the spin-spin susceptibility that can be written in terms of its inverse Fourier amplitude $\chi\left(\textbf{q},\omega\right)$, 
\begin{align*}
\chi(\textbf{r},t)=& \frac{\sqrt{V_0}}{(2\pi)^3}\int d^3q\int\frac{d\omega}{2\pi} e^{i(\textbf{q}\cdot\textbf{r}-\omega t)} \chi\left(\textbf{q},\omega\right).
\end{align*}
where $V_0$ is system volume. We model the metal as a free electron gas with momentum $\hbar\mathbf{q}$ and energy $\epsilon_\textbf{q}= \hbar^2q^2/\left(2m_e\right)$, with $q=\vert \mathbf{q}\vert$ and $m_e$ being the electron mas. The Fermi-Dirac distribution is $f_{\textbf{q}}$. At zero-temperature, $f_{\textbf{q}}= \Theta(k_F-q)=1$ for $q<k_F$ and $f_{\textbf{q}}=0$ for $q>k_F$, $k_F$ being Fermi wave number. For a static impurity spin, the susceptibility reduces to the well-known RKKY function
\begin{align}
\chi_0(r)= \frac{\text{DOS}(k_F)}{4\pi r^3}\left(\frac{\sin 2k_Fr}{2k_Fr}-\cos 2k_Fr\right),
\label{RKKYSimple}
\end{align}
where the density of states is DOS$(k_F)=m_ek_F/\left(\pi\hbar\right)^{2}$. Using the Fourier amplitude $\textbf{S}_1\left(\omega\right)=2\pi\delta\left(\omega\right)\mathbf{S}_1$ of  $\textbf{S}_1$, we get
\begin{align*}
\langle\textbf{s}\rangle_{\rm RKKY}(\textbf{r},t)
=& J_\text{3d}   \chi_0\left(\textbf{r}-\textbf{R}_1\right) \int \frac{d\omega}{2\pi} e^{-i\omega t}  \textbf{S}_1\left(\omega\right)=J_\text{3d}   \chi_0\left(\textbf{r}-\textbf{R}_1\right)  \textbf{S}_1.
\end{align*}
The spin density $\langle\textbf{s}\rangle_{\rm RKKY}(\textbf{r},t)$ mediates the isotropic RKKY interaction between $\mathbf{S}_1$ and a second spin $\mathbf{S}_2$ at $\mathbf{R}_1$,
\begin{align*}
H_\mathrm{RKKY} = - J_\text{3d} ^2 \textbf{S}_1\cdot \textbf{S}_2 \chi (\textbf{R}_2-\textbf{R}_1) .
\end{align*}
\section{Anisotropic exchange due to rare-earth impurity}
When there is a rare-earth impurity at $\mathbf{r}=0$, we consider the spin-spin exchange ($H_{\rm Ss}$) and the spin-dependent skew-scattering ($H_{SsLl} $) Hamiltonians, which in their second-quantization form read
\begin{align*}
H_{\rm Ss}=& -\frac{J_\text{3d}\hbar}{2V_0}\sum_{\textbf{k}_1\textbf{k}_2\alpha'\beta'}  a^\dagger_{\textbf{k}_1\alpha'} a_{\textbf{k}_2\beta'}  \boldsymbol{\sigma}_{\alpha'\beta'}\cdot\textbf{S}_{1},\\
H_{SsLl}
=& -i\frac{\xi}{V_0}\sum_{\textbf{k}_1\textbf{k}_2\alpha'\beta'}  a^\dagger_{\textbf{k}_1\alpha'} a_{\textbf{k}_2\beta'}  \left(\boldsymbol{\sigma}_{\alpha'\beta'}\cdot\textbf{S}_{\rm f} \right)\left(\textbf{L}_{\rm f}\cdot \left[\textbf{k}_1\times\textbf{k}_2\right]\right),
\end{align*}
where the rare-earth orbital momentum $\mathbf{L}_{\rm f}=\left(2-g_J\right)\mathbf{J}_{\rm f}$ is projected onto the total angular momentum $\mathbf{J}_{\rm f}=\mathbf{S}_{\rm f}+\mathbf{L}_{\rm f}$ using the Land\'e g-factor $g_J$, and the spin reads $\mathbf{S}_{\rm f}=\left(g_J-1\right)\mathbf{J}_{\rm f}$. We treat $\mathbf{S}_{\rm f}$, $\mathbf{L}_{\rm f}$, and $\mathbf{J}_{\rm f}$ as classical vectors. The spin-spin interaction between the rare-earth impurity and conduction electrons is described by $H_{\rm Ss}$ after replacing the strength of the s-d interaction ($J_{\rm 3d}$) by the one of the s-f coupling between conduction electrons and 4f electrons ($J_{\rm 4f}$).
The symbol $\xi$ stands for the spin-dependent skew scattering constant, where $\xi\hbar^2k_F^2$ has units of energy multiplied by volume, and $\xi>0$ and $\xi<0$ for more and less than half-filled 4f subshells, respectively. The fermionic operator $a_{\textbf{k}\alpha}$ destroys a state with momentum $\hbar\mathbf{k}$ and spin label $\alpha$. When $H_{\rm Ss}=H_{SsLl}=0$, the total Hamiltonian is equal to the unperturbed one, $H_0=\sum_{\textbf{k},\beta}\epsilon_{\textbf{k}}a_{\textbf{k}\beta}^{\dagger
}a_{\textbf{k}\beta} $, and therefore the time dependence of the destruction operator reads $
a_{\mathbf{k}\alpha}\left(t\right)=e^{-i\epsilon_\mathbf{k}t/\hbar}a_{\mathbf{k}\alpha}\left(0\right)$ and the ensemble mean value of $a^\dagger_{\textbf{k}_1\alpha} a_{\textbf{k}_2\beta}$ is $f_{\mathbf{k_1}}\delta_{\mathbf{k_1},\mathbf{k_2}}\delta_{\alpha,\beta}$.
In second quantization, the spin density reads
 \begin{align*}
\mathbf{s}\left(\mathbf{r}\right)=\frac{1}{V_0}\sum_{\mathbf{p},\mathbf{q}}e^{i\mathbf{r}\cdot\left(\mathbf{q-p}\right)}\sum_{\alpha,\beta}a_{\textbf{p}\alpha}^{\dagger}\left[\frac{\hbar\boldsymbol{\sigma}_{\alpha,\beta}}{2}\right]a_{\textbf{q}\beta}.
\end{align*}
Due to $H_{\rm Ss}$ and $H_{SsLl}$, the presence of a rare-earth atom modifies the ensemble mean value of the spin polarization. Furthermore, the conduction-electron spin density produced by a localized spin $\mathbf{S}_1$ rotates and acquires a component along the $\mathbf{S}_1\times\mathbf{L}_{\rm f}$ direction.

In order to find the ensemble expectation value of the spin-density, $\langle\mathbf{s}\rangle(\textbf{r},t)$, we use the time-evolution operator 
\begin{equation}
U(t)=\exp\left(-\frac{i}{\hbar}\int_{-\infty}^t\left[H_{\rm Ss}\left(t'\right)+H_{SsLl}\left(t'\right)\right]dt'
\right),
\end{equation}
and obtain
\begin{align}
\langle\mathbf{s}\rangle(\textbf{r},t)=\left\langle U^{-1}(t)\langle\mathbf{s}\rangle(\textbf{r},t)U(t)\right\rangle_0\approx\langle\mathbf{s}\rangle_{\rm RKKY}(\textbf{r},t)+\langle\mathbf{s}\rangle_{\rm DMI}(\textbf{r},t)+\mathcal{O}\left(J_{\rm 3d}^2\right),
\label{EqSpinPolIntree}
\end{align}
where $\langle A\rangle_0$ stands for the unperturbed ensemble mean value of $A$, that is, the mean value when $H_{\rm Ss}=H_{SsLl}=0$. The spin polarization induced by the spin-spin exchange and tilted by the spin-orbit coupling is
\begin{align}
\langle\mathbf{s}\rangle_{\rm DMI}(\textbf{r},t)=&-\frac{1}{2\hbar^2}
\left\langle\left[\int_{-\infty}^t H_{\rm Ss}(t^{\prime})dt^{\prime},\left[\int_{-\infty}^t  H_{SsLl}(t^{\prime\prime})dt^{\prime\prime},\mathbf{s}(\textbf{r},t)\right]\right]\right\rangle_0\nonumber\\
&-\frac{1}{2\hbar^2}
\left\langle\left[\int_{-\infty}^t  H_{SsLl}(t^{\prime})dt^{\prime},\left[\int_{-\infty}^t H_{\rm Ss}(t^{\prime\prime})dt^{\prime\prime},\mathbf{s}(\textbf{r},t)\right]\right]\right\rangle_0.
\label{EqStartingPointNonLinear}
\end{align}
Finally, in Eq.~(\ref{EqSpinPolIntree}), the term of order $J_{\rm 3d}^2$ does not provide anisotropic spin polarizations and therefore is a small quantitative correction to the RKKY spin polarization. Consequently, we disregard this term. On the other hand,$\langle\mathbf{s}\rangle_{\rm RKKY}$ accounts for the usual spin-polarization around an impurity spin and its calculation is reviewed in the next subsection.

\subsection{Revision of the RKKY spin susceptibility}
\label{SubSecSS}
The average conduction-electron spin density induced by a localized 3d spin ${S}_1$ at $\mathbf{R_1}$ reads
\begin{align}
\langle{s}_i\rangle(\textbf{r},t)=&J_\text{3d}  \sum_j\int d^3{r'}dt' \chi_{ij}(\textbf{r},\textbf{r}',t-t') {S}_1^j(t')\delta(\textbf{r}'-\mathbf{R}_1)= J_\text{3d}  \sum_j\int dt' \chi_{ij}(\textbf{r},\textbf{R}_1,t-t') {S}^j_1(t'),
\end{align}
 where ${S}^j_1$ is the $j-$th Cartesian component of the vector $\mathbf{S_1}$, and
the spin-spin susceptibility is
\begin{align}
\chi_{ij}(\textbf{r}_1,\textbf{r}_2,t)  =&\frac{i}{\hbar}\theta(t)\left\langle
[s_i(\textbf{r}_1,t),s_j(\textbf{r}_2,0)]\right\rangle_0 \nonumber\\
=&\frac{\hbar^2}{4V_0^2}\sum_{\textbf{pq}\textbf{p}'\textbf{q}'}
e^{ i\left[(\textbf{q}-\textbf{p})\cdot \textbf{r}_1+(\textbf{q}'-\textbf{p}')\cdot \textbf{r}_2\right]}
\sum_{\alpha\beta\alpha'\beta'}\frac{i}{\hbar}\theta(t)\left\langle [a_{\textbf{p}\alpha
}^{\dagger}(t)\sigma^i_{\alpha\beta}a_{\textbf{q}\beta}(t),
a_{\textbf{p}'\alpha'
}^{\dagger}(0)\sigma^j_{\alpha'\beta'}a_{\textbf{q}'\beta'}(0)]\right\rangle_0\nonumber \\
\equiv&\frac{\hbar^2}{4V_0^2} \sum_{\textbf{pq}\textbf{p}'\textbf{q}'}
e^{ i\left[(\textbf{q}-\textbf{p})\cdot \textbf{r}_1+(\textbf{q}'-\textbf{p}')\cdot \textbf{r}_2\right]}
\tilde{\chi}_{ij}
(\textbf{p},\textbf{q},\textbf{p}',\textbf{q}',t).\label{SusceptDef1}
\end{align}
The equation of motion for $\tilde{\chi}_{ij}
(\textbf{p},\textbf{q},\textbf{p}',\textbf{q}',t)$ is
\begin{align*}
&i\hbar\frac{\partial}{\partial t}\tilde{\chi}_{ij}
(\textbf{p},\textbf{q},\textbf{p}',\textbf{q}',t)  
=i\frac{\partial}{\partial t}\left(\sum_{\alpha\beta\alpha'\beta'}i\theta(t)\left\langle [a_{\textbf{p}\alpha}^{\dagger}(t)\sigma^i_{\alpha\beta}a_{\textbf{q}\beta}(t),a_{\textbf{p}'\alpha'
}^{\dagger}(0)\sigma^j_{\alpha'\beta'}a_{\textbf{q}'\beta'}(0)]\right\rangle_0\right) \\
&= \sum_{\alpha\beta\alpha'\beta'}\left(-\delta(t)\left\langle [a_{\textbf{p}\alpha}^{\dagger
}(t)\sigma^i_{\alpha\beta}a_{\textbf{q}\beta}(t),a_{\textbf{p}'\alpha'
}^{\dagger}(0)\sigma^j_{\alpha'\beta'}a_{\textbf{q}'\beta'}(0)]\right\rangle_0
  +\frac{i}{\hbar}\theta(t)\left\langle \left[  [a_{\textbf{p}\alpha}^{\dagger}(t)\sigma^i_{\alpha\beta}a_{\textbf{q}\beta}(t),H_0(t)],a_{\textbf{p}'\alpha'
}^{\dagger}(0)\sigma^j_{\alpha'\beta'}a_{\textbf{q}'\beta'}(0)\right]
\right\rangle_0 \right). 
\end{align*}
Using $[a_i^\dagger a_j,a_k^\dagger a_l]= \delta_{jk} a_i^\dagger a_l-\delta_{il} a_k^\dagger a_j,$ one obtains
\begin{align*} 
\sum_{\alpha\beta}[a_{\textbf{p}\alpha}^{\dagger}{\sigma}^i_{\alpha\beta} a_{\textbf{q}\beta},H_{0}]=\sum_{\alpha\beta}\left(  \epsilon_{\textbf{q}}-\epsilon_{\textbf{p}}\right)  a_{\textbf{p}\alpha}^{\dagger}{\sigma}^i_{\alpha\beta} a_{\textbf{q}\beta} 
\end{align*}
and
\begin{align*} 
\frac{2V_0}{J_\text{3d}\hbar}\sum_{\alpha\beta}[a_{\textbf{p}\alpha}^{\dagger}{\sigma}^i_{\alpha\beta} a_{\textbf{q}\beta},H_{\rm Ss}]
=& -\sum_{\alpha\beta}\sum_{k\beta'}\sum_{\textbf{k}_2} a^\dagger_{\textbf{p}\alpha} a_{\textbf{k}_2\beta'} {\sigma}^i_{\alpha\beta} {\sigma}^k_{\beta\beta'}S^k_{1} 
+ \sum_{\alpha\beta}\sum_{k\alpha'}\sum_{\textbf{k}_1} a^\dagger_{\textbf{k}_1\alpha'} a_{\textbf{q}\beta} {\sigma}^i_{\alpha\beta}{\sigma}^k_{\alpha'\alpha}S^k_{1} \\
=& -\sum_{\alpha}\sum_{k\beta'}\sum_{\textbf{k}_2} a^\dagger_{\textbf{p}\alpha} a_{\textbf{k}_2\beta'} \left(\delta_{\alpha\beta'}\delta_{ik}+\sum_li\varepsilon_{ikl}\sigma^l_{\alpha\beta'}\right) S^k_{1} \nonumber\\&
+ \sum_{\beta}\sum_{k\alpha'}\sum_{\textbf{k}_1} a^\dagger_{\textbf{k}_1\alpha'} a_{\textbf{q}\beta} \left(\delta_{\alpha'\beta}\delta_{ik}+\sum_li\varepsilon_{kil}{\sigma}^l_{\alpha'\beta}\right)S^k_{1} \\
=& \sum_{\alpha\textbf{k}} \left( a^\dagger_{\textbf{k}\alpha} a_{\textbf{q}\alpha} - a^\dagger_{\textbf{p}\alpha} a_{\textbf{k}\alpha}\right)S^i_{1} 
+ i\sum_{kl\alpha\beta}\varepsilon_{ilk}\sigma_{\alpha\beta}^lS^k_{1} \sum_{\textbf{k}}\left(   a^\dagger_{\textbf{k}\alpha} a_{\textbf{q}\beta}  + a^\dagger_{\textbf{p}\alpha} a_{\textbf{k}\beta}\right).
\end{align*}
We used then identity $\left(  \mathbf{A}\cdot\boldsymbol{\sigma}\right)  \left(  \mathbf{B}%
\cdot\boldsymbol{\sigma}\right)  =\left(  \mathbf{A}\cdot\mathbf{B}\right)
\mathbb{I}+i\boldsymbol{\sigma}\cdot\left(  \mathbf{A}\times\mathbf{B}\right),$ with $\mathbf{A}$ and $\mathbf{B}$ being three-component vectors and $\mathbb{I}$ is the $2\times2$ identity matrix. Using the same identity,
\begin{align*}
\sum_{\alpha\beta\alpha'\beta'}\left\langle [a_{\textbf{p}\alpha}^{\dagger}\sigma^{i}_{\alpha\beta} a_{\textbf{q}\beta},a_{\textbf{p}'\alpha'}^{\dagger}\sigma^{j}_{\alpha'\beta'} a_{\textbf{q}'\beta'}]\right\rangle_0=&\sum_{\alpha\beta}
\left\langle -\sum_{\alpha'}\delta_{\textbf{p}\textbf{q}'}a_{\textbf{p}'\alpha'}^{\dagger}\sigma^{j}_{\alpha'\alpha} \sigma^{i}_{\alpha\beta} a_{\textbf{q}\beta}
+\sum_{\beta'} \delta_{\textbf{q}\textbf{p}'} a_{\textbf{p}\alpha}^{\dagger}\sigma^{i}_{\alpha\beta}\sigma^{j}_{\beta\beta'} a_{\textbf{q}'\beta'}\right\rangle_0 \\
=& 2\delta_{ij}\delta_{\textbf{q}\textbf{p}'}\delta_{\textbf{p}\textbf{q}'}\left(f_{\textbf{p}
}-f_{\textbf{q}}\right).
\end{align*}
Then,
\begin{align*}
 \sum_{\alpha\beta\alpha'\beta'}\frac{i}{\hbar}\theta(t)&\left\langle \left[  [a_{\textbf{p}\alpha}^{\dagger}(t)\sigma^i_{\alpha\beta}a_{\textbf{q}\beta}(t),H_0(t)],a_{\textbf{p}'\alpha'
}^{\dagger}(0)\sigma^j_{\alpha'\beta'}a_{\textbf{q}'\beta'}(0)\right]
\right\rangle_0,\notag\\=& \frac{i}{\hbar}\theta(t) \sum_{\alpha'\beta'}\left\langle \left[\sum_{\alpha\beta}\left(  \epsilon_{\textbf{q}}-\epsilon_{\textbf{p}}\right)  a_{\textbf{p}\alpha}^{\dagger}(t){\sigma}^i_{\alpha\beta} a_{\textbf{q}\beta} (t),a_{\textbf{p}'\alpha'
}^{\dagger}(0)\sigma^j_{\alpha'\beta'}a_{\textbf{q}'\beta'}(0)\right]\right\rangle_0,\\
=&\left(\epsilon_\mathbf{q}-\epsilon_\mathbf{p}\right)\tilde{\chi}_{ij}(\textbf{p},\textbf{q},\textbf{p}',\textbf{q}',t).
\end{align*}
With the above calculations, the equation of the spin-spin susceptibility is
\begin{align*}
i\hbar\frac{\partial}{\partial t}\tilde{\chi}_{ij}(\textbf{p},\textbf{q},\textbf{p}',\textbf{q}',t)  
=&-2\delta(t)\delta_{ij}\delta_{\textbf{q}\textbf{p}'}\delta_{\textbf{p}\textbf{q}'}\left(  f_{\textbf{p}}-f_{\textbf{q}}\right)-\left(  \epsilon
_{\textbf{p}}-\epsilon_{\textbf{q}}\right)  \tilde{\chi}_{ij}(\textbf{p},\textbf{q},\textbf{p}',\textbf{q}',t).
\end{align*}
Using the Fourier transforms $\chi(\omega)=\int dte^{i\omega t}\chi(t)$ and $\chi(t)=(2\pi)^{-1}\int d\omega e^{-i\omega t}\chi(\omega)$, the retarded response function becomes
\begin{align*}
\tilde{\chi}_{ij}(\textbf{p},\textbf{q},\textbf{p}',\textbf{q}',\omega) = -2\delta_{ij}\delta_{\textbf{q}\textbf{p}'}\delta_{\textbf{p}\textbf{q}'}
\frac{  f_{\textbf{p}}-f_{\textbf{q}}}{ {\hbar\omega+\epsilon}_{\textbf{p}}-{\epsilon}_{\textbf{q}} +i0^+}.
\end{align*}Using \textit{Dirac's formula}
\begin{align}
\int dz \frac{f(z)}{z-z_0+i0^+}&=\mathcal{P}\left[\int dz \frac{f(z)}{z-z_0}\right]-i\pi f(z_0),\label{DiracsFormula}
\end{align}
where $\mathcal{P}$ stands for the principal part of the integral. Considering that the equilibrium response (i.e., static susceptibility) is a real-valued function, the static susceptibility reads
\begin{align}
\boxed{
\chi_{ij}(\textbf{r}_1,\textbf{r}_2)  =-\frac{\hbar^2\delta_{ij}}{2V_0^2} \mathcal{P}
\sum_{\textbf{pq}}e^{i(\textbf{q}-\textbf{p})\cdot \left(\textbf{r}_1-\textbf{r}_2\right)}\frac{  f_{\textbf{p}}-f_{\textbf{q}}}{ {\epsilon}_{\textbf{p}}-{\epsilon}_{\textbf{q}}}}
\label{EqSusc}
\end{align}%
which becomes the RKKY function, Eq.~(\ref{RKKYSimple}), after integration. Note that, at leading order in the $J_{\rm 3d}$ parameter, the susceptibility $\chi_{ij}(\textbf{r}_1,\textbf{r}_2)$ depends only on $\textbf{r}_1-\textbf{r}_2$. This is because the nonuniformity induced by the rare-earth atom is a higher-order correction, i.e., the terms that break the space-translation invariance scale as $\left(J_{\rm ex}\right)^2$ or $J_{\rm ex}\xi$.

\section{Dzyaloshinskii-Moriya spin-susceptibility}
\label{SubSecSkew}
Let us write the Dzyaloshinskii-Moriya spin density as
\begin{align}
\langle\mathbf{s}\rangle_{\rm DMI}(\textbf{r},t)=&\sum_{jk} \frac{i\xi J_{\rm 3d}}{\hbar} \int_{-\infty}^{\infty} dt'\int_{-\infty}^{\infty} dt'' \left[\phi_{kji}(\textbf{r},t'',t',t)S_{1}^{k}(t')S_{f}^{j}(t'')+\varphi_{kji}(\textbf{r},t'',t',t)S_{f}^{k}(t')S_{1}^{j}(t'')\right]\mathbf{e_i}
\label{EqStartingSkew}
\end{align}
where $\mathbf{e_i}$ is the $i-$th Cartesian unit vector. Defining $\mathbf{s}\left(\mathbf{p},\textbf{q},t\right)=\sum_{\alpha,\beta}a_{\textbf{p}\alpha}^{\dagger}\left(t\right)\left[\hbar\boldsymbol{\sigma}_{\alpha,\beta}/2\right]a_{\textbf{q}\beta}\left(t\right)$, the response functions are
\begin{align}
\phi_{kji}(\textbf{r},t'',t',t)=&-\frac{\theta(t-t')\theta(t-t'')}{\hbar^2V_0^2}\sum_{\textbf{p}'\textbf{q}'\textbf{p}''\textbf{q}''}e^{i\textbf{R}_1\cdot(\textbf{q}'-\textbf{p}')}(\mathbf{L_f}(t'')\cdot[\textbf{p}''\times\textbf{q}''])\left\langle\left[ {s}_k(\textbf{p}',\textbf{q}',t'), \left[ s_j(\textbf{p}'',\textbf{q}'',t''), s_i(\textbf{r},t)\right]\right]\right\rangle_0\notag\\
\varphi_{kji}(\textbf{r},t'',t',t)=&-\frac{\theta(t-t')\theta(t-t'')}{\hbar^2V_0^2}\sum_{\textbf{p}'\textbf{q}'\textbf{p}''\textbf{q}''}e^{i\textbf{R}_1\cdot(\textbf{q}''-\textbf{p}'')}(\mathbf{L_f}(t')\cdot[\textbf{p}'\times\textbf{q}'])\left\langle\left[ {s}_k(\textbf{p}',\textbf{q}',t'), \left[ s_j(\textbf{p}'',\textbf{q}'',t''), s_i(\textbf{r},t)\right]\right]\right\rangle_0.
\end{align}
For both $\phi_{kji}$ and $\varphi_{kji}$, we need to obtain 
\begin{align*}
A=&\sum_{\alpha''\beta''}[a_{\textbf{p}''\alpha''}^{\dagger}(t''){\sigma}^j_{\alpha''\beta''} a_{\textbf{q}''\beta''}(t''),s_i(\textbf{r},t)]\\
=&\frac{\hbar}{2V_0}\sum_{\alpha\beta} e^{\frac{i}{\hbar}\left(\epsilon_{\mathbf p''}-\epsilon_{\mathbf q''}\right)t''}\left(\sum_{\textbf{q}}e^{\frac{i}{\hbar}\left(\epsilon_{\mathbf q''}-\epsilon_{\mathbf q}\right)t}e^{i(\textbf{q}-\textbf{q}'')\cdot\textbf{r}}\left[\delta_{ij}\delta_{\alpha\beta}-i\sum_l\epsilon_{ijl}\sigma^l_{\alpha\beta} \right]a_{\textbf{p}''\alpha}^{\dagger}(0)a_{\textbf{q}\beta}(0)\notag \right.\\
&\left. -\sum_{\textbf{p}}e^{\frac{i}{\hbar}\left(\epsilon_{\mathbf p}-\epsilon_{\textbf{p}''}\right)t}e^{i(\textbf{p}''-\textbf{p})\cdot\textbf{r}}\left[\delta_{ij}\delta_{\alpha\beta}+i\sum_l\epsilon_{ijl}\sigma^l_{\alpha\beta}\right]a_{\textbf{p}\alpha}^{\dagger}(0)a_{\textbf{q}''\beta}(0)\right)
\end{align*} 
and
\begin{align*}
B&=\left<\sum_{\alpha'\beta'}[a_{\textbf{p}'\alpha'}^{\dagger}(t'){\sigma}^k_{\alpha'\beta'} a_{\textbf{q}'\beta'}(t'),A]\right>_0\\
=&-i\epsilon_{ijk}\frac{\hbar}{V_0}\left(f_{\textbf{p}'}-f_{\textbf{q}'}\right) e^{\frac{i}{\hbar}\left(\epsilon_{\mathbf p''}-\epsilon_{\mathbf q''}\right)t''}e^{\frac{i}{\hbar}\left(\epsilon_{\mathbf p'}-\epsilon_{\mathbf q'}\right)t'}\left(e^{\frac{i}{\hbar}\left(\epsilon_{\mathbf q''}-\epsilon_{\mathbf p'}\right)t}e^{i(\textbf{p}'-\textbf{q}'')\cdot\textbf{r}}\delta_{\textbf{q}'\textbf{p}''} +e^{\frac{i}{\hbar}\left(\epsilon_{\mathbf q'}-\epsilon_{\textbf{p}''}\right)t}e^{i(\textbf{p}''-\textbf{q}')\cdot\textbf{r}}\delta_{\textbf{p}'\textbf{q}''}\right).
\end{align*}

Therefore, the response function $\phi_{kji}$ reads
\begin{align*}
&\phi_{kji}(\textbf{r},t'',t',t)=-\frac{\theta(t-t')\theta(t-t'')}{\hbar^2V_0^2}\sum_{\textbf{p}'\textbf{q}'\textbf{p}''\textbf{q}''}e^{i\textbf{R}_1\cdot(\textbf{q}'-\textbf{p}')}(\mathbf{L_f}(t'')\cdot[\textbf{p}''\times\textbf{q}''])\frac{\hbar^2B}{4},\notag\\
=&i\epsilon_{ijk}\frac{\hbar \theta(t-t')\theta(t-t'')}{4V_0^3}\sum_{\textbf{p}'\textbf{q}'}e^{i\textbf{R}_1\cdot(\textbf{q}'-\textbf{p}')}\left(f_{\textbf{p}'}-f_{\textbf{q}'}\right) e^{\frac{i}{\hbar}\left(\epsilon_{\mathbf p'}-\epsilon_{\mathbf q'}\right)t'}\mathbf{L_f}(t'')\\
& \cdot\left(\sum_{\textbf{q}''}e^{\frac{i}{\hbar}\left(\epsilon_{\mathbf q'}-\epsilon_{\mathbf q''}\right)t''}\textbf{q}'\times\textbf{q}''e^{\frac{i}{\hbar}\left(\epsilon_{\mathbf q''}-\epsilon_{\mathbf p'}\right)t}e^{i(\textbf{p}'-\textbf{q}'')\cdot\textbf{r}}
+\sum_{\textbf{p}''}e^{\frac{i}{\hbar}\left(\epsilon_{\mathbf p''}-\epsilon_{\mathbf p'}\right)t''}\textbf{p}''\times\textbf{p}'e^{\frac{i}{\hbar}\left(\epsilon_{\mathbf q'}-\epsilon_{\textbf{p}''}\right)t}e^{i(\textbf{p}''-\textbf{q}')\cdot\textbf{r}} \right),
\end{align*}
and $\varphi_{kji}$ is
\begin{align*}
&\varphi_{kji}(\textbf{r},t'',t',t)=-\frac{\theta(t-t')\theta(t-t'')}{\hbar^2V_0^2}\sum_{\textbf{p}'\textbf{q}'\textbf{p}''\textbf{q}''}e^{i\textbf{R}_1\cdot(\textbf{q}''-\textbf{p}'')}(\mathbf{L_f}(t')\cdot[\textbf{p}'\times\textbf{q}'])\frac{\hbar^2B}{4},\notag\\
=&i\epsilon_{ijk}\frac{\hbar \theta(t-t')\theta(t-t'')}{4 V_0^3}\sum_{\textbf{p}'\textbf{q}'}\left(f_{\textbf{p}'}-f_{\textbf{q}'}\right) e^{\frac{i}{\hbar}\left(\epsilon_{\mathbf p'}-\epsilon_{\mathbf q'}\right)t'}\mathbf{L_f}(t')\cdot\left[\textbf{p}'\times\textbf{q}'\right]\\
& \cdot\left(\sum_{\textbf{q}''}e^{i\textbf{R}_1\cdot(\textbf{q}''-\textbf{q}')}e^{\frac{i}{\hbar}\left(\epsilon_{\mathbf q'}-\epsilon_{\mathbf q''}\right)t''}e^{\frac{i}{\hbar}\left(\epsilon_{\mathbf q''}-\epsilon_{\mathbf p'}\right)t}e^{i(\textbf{p}'-\textbf{q}'')\cdot\textbf{r}} +\sum_{\textbf{p}''}e^{i\textbf{R}_1\cdot(\textbf{p}'-\textbf{p}'')}e^{\frac{i}{\hbar}\left(\epsilon_{\mathbf p''}-\epsilon_{\mathbf p'}\right)t''}e^{\frac{i}{\hbar}\left(\epsilon_{\mathbf q'}-\epsilon_{\textbf{p}''}\right)t}e^{i(\textbf{p}''-\textbf{q}')\cdot\textbf{r}}\right).
\end{align*}
Using direct and inverse Fourier transforms,
\begin{align*}
\theta(t)=\int \frac{d\omega}{2\pi} \frac{e^{-i\omega t}}{0^+-i\omega},
\end{align*}
and a static orbital momentum, $\mathbf{L_{\rm}}(t)=\mathbf{L_{\rm}}$, we arrive at
\begin{align*}
&\phi_{kji}(\textbf{r},t'',t',t)=i\epsilon_{ijk}\frac{\hbar}{4 V_0^3}\iint \frac{d\omega'd\omega''}{(2\pi)^2}\frac{e^{-i\omega't}}{0^+-i\omega'}\frac{e^{-i\omega''t}}{0^+-i\omega''}\sum_{\textbf{p}'\textbf{q}'}e^{i\textbf{R}_1\cdot(\textbf{q}'-\textbf{p}')}\left(f_{\textbf{p}'}-f_{\textbf{q}'}\right) e^{\frac{i}{\hbar}\left(\epsilon_{\mathbf p'}-\epsilon_{\mathbf q'}+\hbar\omega'\right)t'}    \mathbf{L_{\rm}}\\
& \cdot\left(\sum_{\textbf{q}''}e^{\frac{i}{\hbar}\left(\epsilon_{\mathbf q'}-\epsilon_{\mathbf q''}+\hbar\omega''\right)t''}\textbf{q}'\times\textbf{q}'' e^{\frac{i}{\hbar}\left(\epsilon_{\mathbf q''}-\epsilon_{\mathbf p'}\right)t}e^{i(\textbf{p}'-\textbf{q}'')\cdot\textbf{r}}
+\sum_{\textbf{p}''}e^{\frac{i}{\hbar}\left(\epsilon_{\mathbf p''}-\epsilon_{\mathbf p'}+\hbar\omega''\right)t''}\textbf{p}''\times\textbf{p}' e^{\frac{i}{\hbar}\left(\epsilon_{\mathbf q'}-\epsilon_{\mathbf p''}\right)t}e^{i(\textbf{p}''-\textbf{q}')\cdot\textbf{r}} \right).
\end{align*}

\begin{align*}
&\varphi_{kji}(\textbf{r},t'',t',t)=i\epsilon_{ijk}\frac{\hbar}{4 V_0^3}\iint \frac{d\omega'd\omega''}{(2\pi)^2}\frac{e^{-i\omega't}}{0^+-i\omega'}\frac{e^{-i\omega''t}}{0^+-i\omega''}\sum_{\textbf{p}'\textbf{q}'}\left(f_{\textbf{p}'}-f_{\textbf{q}'}\right) e^{\frac{i}{\hbar}\left(\epsilon_{\mathbf p'}-\epsilon_{\mathbf q'}+\hbar\omega'\right)t'}\mathbf{L_f}\cdot\left[\textbf{p}'\times\textbf{q}'\right]\\
& \cdot\left(\sum_{\textbf{q}''}e^{i\textbf{R}_1\cdot(\textbf{q}''-\textbf{q}')}e^{\frac{i}{\hbar}\left(\epsilon_{\mathbf q'}-\epsilon_{\mathbf q''}+\hbar\omega''\right)t''}e^{\frac{i}{\hbar}\left(\epsilon_{\mathbf q''}-\epsilon_{\mathbf p'}\right)t}e^{i(\textbf{p}'-\textbf{q}'')\cdot\textbf{r}} +\sum_{\textbf{p}''}e^{i\textbf{R}_1\cdot(\textbf{p}'-\textbf{p}'')}e^{\frac{i}{\hbar}\left(\epsilon_{\mathbf p''}-\epsilon_{\mathbf p'}+\hbar\omega''\right)t''}e^{\frac{i}{\hbar}\left(\epsilon_{\mathbf q'}-\epsilon_{\textbf{p}''}\right)t}e^{i(\textbf{p}''-\textbf{q}')\cdot\textbf{r}}\right).
\end{align*}
From this point, we consider static spins, $\mathbf{S}_{\rm f}(t)=\mathbf{S}_{\rm f}$ and $\mathbf{S}_1(t)=\mathbf{S}_1$. Integrating over $t'$, $t''$, and 
writing the spin density in terms of the contributions arising from the $\phi$ and $\varphi$ response functions, $\textbf{s}(\textbf{r})=\textbf{s}^\phi(\textbf{r})+\textbf{s}^\varphi(\textbf{r})$ can be simplified by appropriately renaming $\textbf{p}',\textbf{q}',\textbf{p}'',\textbf{q}''$

\begin{align*}
&s_{ijk}^\phi(\textbf{r})=\frac{i\xi J_{\rm 3d}}{\hbar}\iint  dt' dt'' \phi_{kji}(\textbf{r},t'',t',t) S_1^kS_f^j,\\
=&\epsilon_{ijk}\frac{\xi J_{\rm 3d}\hbar^2S_1^kS_f^j}{4 V_0^3}\sum_{\textbf{kpq}}f_{\textbf{k}} \frac{1}{\epsilon_{\mathbf k}-\epsilon_{\mathbf q}-i0^+}\mathbf{L_f}
 \cdot\left(\frac{\textbf{q}\times\textbf{p} e^{i\left(\textbf{k}\cdot[\textbf{r}-\textbf{R}_1] -\textbf{p}\cdot\textbf{r}+\textbf{q}\cdot\textbf{R}_1\right)}}{\epsilon_{\mathbf q}-\epsilon_{\mathbf p}-i0^+}
+\frac{\textbf{p}\times\textbf{k} e^{i\left(-\textbf{k}\cdot\textbf{R}_1 +\textbf{p}\cdot\textbf{r}-\textbf{q}\cdot[\textbf{r}-\textbf{R}_1]\right)}}{\epsilon_{\mathbf p}-\epsilon_{\mathbf k}-i0^+} \right)\\
&-\epsilon_{ijk}\frac{\xi J_{\rm 3d}\hbar^2S_1^kS_f^j}{4 V_0^3}\sum_{\textbf{kpq}}f_{\textbf{k}} \frac{1}{\epsilon_{\mathbf q}-\epsilon_{\mathbf k}-i0^+}\mathbf{L_f}\cdot\left(\frac{\textbf{k}\times\textbf{p} e^{i\left(\textbf{k}\cdot\textbf{R}_1 -\textbf{p}\cdot\textbf{r}+\textbf{q}\cdot[\textbf{r}-\textbf{R}_1]\right)}}{\epsilon_{\mathbf k}-\epsilon_{\mathbf p}-i0^+}
+\frac{\textbf{p}\times\textbf{q} e^{i\left(-\textbf{k}\cdot[\textbf{r}-\textbf{R}_1] +\textbf{p}\cdot\textbf{r}-\textbf{q}\cdot\textbf{R}_1\right)}}{\epsilon_{\mathbf p}-\epsilon_{\mathbf q}-i0^+}\right)
\end{align*}
and
\begin{align*}
&s_{ijk}^\varphi(\textbf{r})=\frac{i\xi J_{\rm 3d}}{\hbar}\iint  dt' dt'' \varphi_{kji}(\textbf{r},t'',t',t) S_f^kS_1^j,\\
=&\epsilon_{ijk}\frac{\xi J_{\rm 3d}S_f^kS_1^j\hbar^2}{4 V_0^3}\sum_{\textbf{k}\textbf{q}}f_{\textbf{k}} \frac{\mathbf{L_f}\cdot\left[\textbf{k}\times\textbf{q}\right]}{\epsilon_{\mathbf k}-\epsilon_{\mathbf q}-i0^+}\left(\sum_{\textbf{p}} \frac{e^{i\left(\textbf{k}\cdot \textbf{r}-\textbf{p}\cdot[\textbf{r}-\textbf{R}_1]-\textbf{q}\cdot\textbf{R}_1\right)} }{\epsilon_{\mathbf q}-\epsilon_{\mathbf p}-i0^+}  
+\sum_{\textbf{p}}\frac{ e^{i\left(\textbf{k}\cdot \textbf{R}_1+\textbf{p}\cdot[\textbf{r}-\textbf{R}_1]-\textbf{q}\cdot\textbf{r}\right)} }{\epsilon_{\mathbf p}-\epsilon_{\mathbf k}-i0^+} \right)\\
&-\epsilon_{ijk}\frac{\xi J_{\rm 3d}S_f^kS_1^j\hbar^2}{4 V_0^3}\sum_{\textbf{q}\textbf{k}}f_{\textbf{k}} \frac{\mathbf{L_f}\cdot\left[\textbf{q}\times\textbf{k}\right]}{\epsilon_{\mathbf q}-\epsilon_{\mathbf k}-i0^+}\left(\sum_{\textbf{p}} \frac{ e^{i\left(-\textbf{k}\cdot \textbf{R}_1-\textbf{p}\cdot[\textbf{r}-\textbf{R}_1]+\textbf{q}\cdot\textbf{r}\right)} }{\epsilon_{\mathbf k}-\epsilon_{\mathbf p}-i0^+} 
+\sum_{\textbf{p}}\frac{ e^{i\left(-\textbf{k}\cdot \textbf{r}+\textbf{p}\cdot[\textbf{r}-\textbf{R}_1]+\textbf{q}\cdot\textbf{R}_1\right)} }{\epsilon_{\mathbf p}-\epsilon_{\mathbf q}-i0^+}\right)\\
\end{align*}
setting $\textbf{r}_1=\textbf{r}-\textbf{R}_1$
\begin{align*}
&\textbf{s}^\phi(\textbf{r})=\sum_{ijk}s_{ijk}^\phi(\textbf{r})\mathbf{e_i}\\
=&\textbf{S}_{\rm f}\times\textbf{S}_1\frac{\xi J_{\rm 3d}\hbar^2}{2 V_0^3} \mathrm{Re}\sum_{\textbf{kpq}}f_{\textbf{k}} \frac{\mathbf{L_f}\cdot\left(\nabla_{\textbf{R}_1}\times\nabla_{\textbf{r}}\right) }{\epsilon_{\mathbf k}-\epsilon_{\mathbf q}-i0^+} \left(\frac{e^{i\left(\textbf{k}\cdot\textbf{r}_1 -\textbf{p}\cdot\textbf{r}+\textbf{q}\cdot\textbf{R}_1\right)}}{\epsilon_{\mathbf q}-\epsilon_{\mathbf p}-i0^+}
-\frac{e^{i\left(-\textbf{k}\cdot\textbf{R}_1 +\textbf{p}\cdot\textbf{r}-\textbf{q}\cdot\textbf{r}_1\right)}}{\epsilon_{\mathbf p}-\epsilon_{\mathbf k}-i0^+} \right)
\end{align*}

\begin{align*}
&\textbf{s}^\varphi(\textbf{r})=\sum_{ijk}s_{ijk}^\varphi(\textbf{r})\mathbf{e_i}\\
=&\textbf{S}_1\times\textbf{S}_{\rm f}\frac{\xi J_{\rm 3d}\hbar^2}{2 V_0^3}\mathrm{Re} \sum_{\textbf{kpq}}f_{\textbf{k}} \frac{\mathbf{L_f}\cdot\left[\nabla_\textbf{r}\times\nabla_{\textbf{R}_1}\right]}{\epsilon_{\mathbf k}-\epsilon_{\mathbf q}-i0^+}\left( \frac{e^{i\left(\textbf{k}\cdot \textbf{r}-\textbf{p}\cdot\textbf{r}_1-\textbf{q}\cdot\textbf{R}_1\right)} }{\epsilon_{\mathbf q}-\epsilon_{\mathbf p}-i0^+}  
-\frac{ e^{i\left(\textbf{k}\cdot \textbf{R}_1+\textbf{p}\cdot\textbf{r}_1-\textbf{q}\cdot\textbf{r}\right)} }{\epsilon_{\mathbf p}-\epsilon_{\mathbf k}-i0^+} \right)
\end{align*}

Therefore
\begin{align*}
&\textbf{s}(\textbf{r})=\textbf{S}_{\rm f}\times\textbf{S}_1\frac{\xi J_{\rm 3d}\hbar^2}{ 2V_0^3} \mathrm{Re}\sum_{\textbf{kpq}}f_{\textbf{k}} \frac{\mathbf{L_f}\cdot\left(\nabla_{\textbf{R}_1}\times\nabla_{\textbf{r}}\right) }{\epsilon_{\mathbf k}-\epsilon_{\mathbf q}-i0^+}\notag\\&\cdot \left(\frac{e^{i\left(\textbf{k}\cdot\textbf{r}_1 -\textbf{p}\cdot\textbf{r}+\textbf{q}\cdot\textbf{R}_1\right)}}{\epsilon_{\mathbf q}-\epsilon_{\mathbf p}-i0^+}
-\frac{e^{i\left(-\textbf{k}\cdot\textbf{R}_1 +\textbf{p}\cdot\textbf{r}-\textbf{q}\cdot\textbf{r}_1\right)}}{\epsilon_{\mathbf p}-\epsilon_{\mathbf k}-i0^+} 
+ \frac{e^{i\left(\textbf{k}\cdot \textbf{r}-\textbf{p}\cdot\textbf{r}_1-\textbf{q}\cdot\textbf{R}_1\right)} }{\epsilon_{\mathbf q}-\epsilon_{\mathbf p}-i0^+}  
-\frac{ e^{i\left(\textbf{k}\cdot \textbf{R}_1+\textbf{p}\cdot\textbf{r}_1-\textbf{q}\cdot\textbf{r}\right)} }{\epsilon_{\mathbf p}-\epsilon_{\mathbf k}-i0^+} \right)
\end{align*}

Using the following summations over $\mathbf p$ and $\mathbf q$ 

\begin{align*}
\frac{1}{V^2_0}\sum_\textbf{ q p}&\frac{e^{i\textbf{q}\cdot\textbf{r}_1-i\textbf{p}\cdot\textbf{r}_2}}{(\epsilon_{\mathbf k}-\epsilon_{\mathbf q}-i0^+)(\epsilon_{\mathbf q}-\epsilon_{\mathbf p}-i0^+)}
=\frac{4m_e^2}{\hbar^4} \frac{1}{(2\pi)^3r_1r_2} \int_0^\infty  p dp \sin pr_2  \left( \frac{e^{-ikr_1}-e^{ipr_1}}{(p+k)(p-k)}\right),
\end{align*}
its real part is
\begin{align*}
&\frac{4m_e^2}{\hbar^4} \frac{1}{(2\pi)^3r_1r_2} \int_0^\infty  p dp \sin pr_2  \left( \frac{\cos kr_1-\cos pr_1}{(p+k)(p-k)}\right)\\
=&\frac{4m_e^2}{\hbar^4} \frac{1}{(2\pi)^3r_1r_2}  \frac{\pi}{2} \left(\frac{e^{i kr_2}\cos kr_1-\frac{1}{2}\left(e^{ik(r_2+r_1)}+e^{i k(r_2-r_1)}\right)}{2}+ \frac{e^{-i kr_2}\cos kr_1-\frac{1}{2}\left(e^{-ik(r_2+r_1)}+e^{-i k(r_2-r_1)}\right)}{2}\right)=0
\end{align*}
In a similar way, to obtain \[\frac{1}{V_0^2}\sum_\textbf{qp}\frac{e^{i\textbf{q}\cdot\textbf{r}_1+i\textbf{p}\cdot\textbf{r}_2}}{(\epsilon_{\mathbf k}-\epsilon_{\mathbf q}-i0^+)(\epsilon_{\mathbf p}-\epsilon_{\mathbf k}-i0^+)},\]we employ the following integrals
\begin{align*}
\frac{1}{V_0}\sum_\textbf{p}\frac{e^{i\textbf{p}\cdot\textbf{r}_2}}{\epsilon_{\mathbf p}-\epsilon_{\mathbf k}-i0^+}
=& \frac{2m_e}{\hbar^2}\frac{e^{ikr_2}}{4\pi r_2},
\end{align*}
and
\begin{align*}
\frac{1}{V_0}\sum_\textbf{q}\frac{e^{i\textbf{q}\cdot\textbf{r}_1}}{\epsilon_{\mathbf k}-\epsilon_{\mathbf q}-i0^+}
=& \frac{2m_e}{\hbar^2}\frac{e^{-ikr_1}}{4\pi r_1}.
\end{align*}
Collecting terms, we get
\begin{align*}
\textbf{s}(\textbf{r})
=&-\textbf{S}_{\rm f}\times\textbf{S}_1\xi J_{\rm 3d} \mathbf{L_f}\cdot\left(\nabla_{\textbf{R}_1}\times\nabla_{\textbf{r}}\right) \frac{m_e^2}{4\pi^2\hbar^2}\int_0^{k_F} \frac{4\pi k dk}{(2\pi)^3}\frac{\sin kR_1 \cos k(r-r_1)}{R_1rr_1}\\
\equiv& -\textbf{S}_{\rm f}\times\textbf{S}_1\xi J_{\rm 3d} \mathbf{L_f}\cdot\left(\hat{\textbf{R}}_1\times\hat{\textbf{r}}\right) \frac{m_e^2}{\left(2\pi \right)^4\hbar^2} \frac{F \left(R_1,r,\left|\textbf{r}-\textbf{R}_1\right|\right)}{R_1r\left|\textbf{r}-\textbf{R}_1\right|}
\end{align*}
where 
\begin{align*}
F\left(R_1,r,r1\right)&=
-\frac{3 \left(k_F^2 (r-r_1+R_1)^2-2\right) \sin (k_F (r-r_1+R_1))}{(r-r_1+R_1)^4}-\frac{3 \left(k_F^2 (-r+r_1+R_1)^2-2\right) \sin (k_F (r-r_1-R_1))}{(-r+r_1+R_1)^4}\\&+\frac{k_F \left(k_F^2 (-r+r_1+R_1)^2-6\right) \cos (k_F (r-r_1-R_1))}{(r-r_1-R_1)^3}+\frac{k_F \left(k_F^2 (r-r_1+R_1)^2-6\right) \cos (k_F (r-r_1+R_1))}{(r-r_1+R_1)^3}.
\end{align*}
Therefore, the Dzyaloshinskii–Moriya interaction strength is
\begin{align}
H=&-J_{\rm 3d} \textbf{S}_2\cdot\langle\mathbf{s}\rangle(\textbf{S}_1,\textbf{R}_2)-J_{\rm 3d} \textbf{S}_1\cdot\langle\mathbf{s}\rangle(\textbf{S}_2,\textbf{R}_1)\notag\\
&+\textbf{S}_1\cdot(\textbf{S}_{\rm f}\times\textbf{S}_2)\xi J^2_{\rm 3d} \mathbf{L_f}\cdot\left(\hat{\textbf{R}}_2\times\hat{\textbf{R}}_1\right)  \frac{m_e^2}{\left(2\pi \right)^4\hbar^2} \frac{F \left(R_2,R_1,\left|\textbf{R}_1-\textbf{R}_2\right|\right)}{R_1R_2\left|\textbf{R}_1-\textbf{R}_2\right|}\notag\\
=&\left(\textbf{S}_1\times\textbf{S}_2\right)\cdot\textbf{S}_{\rm f} \mathbf{L_f}\cdot\left(\hat{\textbf{R}}_1\times\hat{\textbf{R}}_2\right) \frac{\xi J^2_{\rm 3d}m_e^2}{\left(2\pi \right)^4\hbar^2} \frac{F \left(R_1,R_2,\left|\textbf{R}_2-\textbf{R}_1\right|\right)+ F \left(R_2,R_1,\left|\textbf{R}_1-\textbf{R}_2\right|\right)}{R_1R_2\left|\textbf{R}_1-\textbf{R}_2\right|}\notag\\
\equiv&\textbf{D}^f_{12}\cdot\left(\textbf{S}_{1}\times\textbf{S}_{2}\right).
\end{align}
\end{document}